\documentclass[onecolumn,nofootinbib]{revtex4-2}

\usepackage{lmodern}
\usepackage{hyperref}
\usepackage{graphicx}
\usepackage{physics}
\usepackage{amssymb}
\usepackage{bbm}
\usepackage{subfigmat}
\usepackage{xcolor}

\hypersetup{
    colorlinks = true,
    linkcolor  = blue,
    citecolor  = blue,
    urlcolor   = blue,
}

\graphicspath{{Figures/}}

\begin{document}

\title{Quantum retrodiction: foundations and controversies}

\author{Stephen M. Barnett$^{1}$}
\email{stephen.barnett@glasgow.ac.uk}
\author{John Jeffers$^{2}$}
\author{David T. Pegg$^{3}$}

\affiliation{$^{1}$School of Physics and Astronomy, University of 
Glasgow, Glasgow G12 8QQ, U.K.\\$^{2}$Department of Physics, University of Strathclyde, Glasgow G4 0NG, U.K.\\$^{3}$Centre for Quantum Dynamics, School of Science, 
Griffith University, Nathan 4111, Australia}
\date{\today}
\begin{abstract}
Prediction is the making of statements, usually probabilistic, about future events based on current information. Retrodiction is the making of statements about past events based on current information.  We present the foundations of quantum retrodiction and highlight its intimate connection with the Bayesian interpretation of probability.  The close link with  Bayesian methods enables us to explore controversies and misunderstandings about retrodiction that have appeared in the literature.  To be clear, quantum retrodiction is universally applicable and draws its validity directly from conventional predictive quantum theory coupled with Bayes' theorem.\\ \\
Key Words: Quantum foundations; Bayesian inference; Time reversal.
\end{abstract}
\maketitle


\section{Introduction}
\label{Sect1}
Quantum theory is usually presented as a predictive theory, with statements made concerning the probabilities for measurement
outcomes based upon earlier preparation events.  In retrodictive 
quantum theory this order is reversed and we seek to use the outcome 
of a measurement to make probabilistic statements concerning 
\emph{earlier} events \cite{Watanabe,ABL,Belinfante,PB99,BPJ00,SUSSP}.
The theory was first presented within the context of time-reversal
symmetry \cite{Watanabe,ABL,Belinfante} but, more recently, has been 
developed into a practical tool for analysing experiments in quantum optics \cite{PPB98,BPP98,BPJJL00,BPJJ01,TJBP03,Babichev03,Chefles03,TJB04,JLJ06} and other areas such as continuous monitoring \cite{GJM13,RPP15,TWS15,TNK17,BDL20,BJD20,ZM20} and imaging \cite{SJB15}.

The development of retrodictive quantum theory has been accompanied by controversy.  At the root of this lies such profound questions as the nature of probability and the interpretation of the state vector. We review retrodictive quantum theory paying particular attention to the principles that underlie it, drawing on an earlier set of lecture notes \cite{SUSSP}.  We then present the arguments that have been presented to restrict the applicability of quantum retrodiction \footnote{Strictly speaking postdiction, as advocated by Belinfante \cite{Belinfante}, is probably a more accurate term than retrodiction, but retrodiction is the original name for our topic and we stick with it.}, to modify it or to challenge its validity and address the concerns underlying each of these.  Our conclusion is that quantum retrodiction is intimately connected with the Bayesian conception of probability.  Indeed, we can derive quantum retrodiction on the basis of conventional predictive quantum theory plus Bayes' theorem
\footnote{ In the literature one can find the possessive form written as Bayes' and also as Bayes's.  The latter is probably more correct grammatically, but we prefer the former on the grounds that it sounds better.}\cite{BPJ00}
(see also \cite{Rovelli}).  

It is important to realise that our approach is rooted in the Bayesian interpretation of probabilities.  The apparently straightforward decision to adopt a Bayesian perspective commits us, in the quantum theory, to a particular philosophical position on the nature of the state vector or wavefunction.  If we are to retain the notion that probabilities in quantum theory derive from the state vector and also the Bayesian idea that individuals with access to different information assign different probabilities, then in the quantum theory we must also assign different states to a quantum system.  This is at odds with the notion that the wavefunction has any ontological meaning (real existence).  If one starts from the position that the wavefunction does have an ontological meaning, then attempting to develop quantum retrodiction leads to paradoxical conclusions \cite{Peres}.

When viewed within the context of Bayesian inference, we find that quantum retrodiction has general validity and applicability and that it can be employed in any situation to use a measurement result to make probabilistic inferences about earlier, premeasurement occurrences.

We conclude with a few remarks concerning the arrow of time in quantum theory in light of the insights gained from quantum retrodiction.


\section{Predictive and Retrodictive Probabilities}
\label{Sect2}

We begin with a comparison of prediction and retrodiction in a classical setting.  The key to this comparison is Bayes' theorem connecting predictive and retrodictive conditional probabilities. Thus retrodiction is intimately connected with the Bayesian interpretation of probability.

It may come as something of a surprise to the uninitiated that the nature of probability has been in dispute.  There are, however, two distinct schools of thought on this question, the frequentists and the Bayesians.  A frequentist defines a probability in terms of frequency of outcome; if the situation is repeated a great number of times (ideally infinitely many) then the probability for a given outcome is simply the proportion or fraction of times it occurred in the large ensemble.  A Bayesian, however, emphasises information and the state of knowledge, so a probability is a statement of confidence or belief, and can be updated or modified as information is acquired.  The distinction between these views is subtle but important and, indeed, was a source of intense debate for much of the twentieth century.  There is not the space here to go into the arguments and to do so would distract us from our principal topic.  We can get a flavour of the issues, however, from the writings of two of the earlier protagonists. Fisher \cite{Fisher}, for the frequentists, wrote `{\em ... it will be sufficient in this general outline of the scope of Statistical Science to express my personal conviction, which I have sustained elsewhere, that the theory of inverse probability [Bayesian Inference] is founded upon an error and must be wholly rejected.}'.  Much the same view seems to have been expressed by Gibbs, in the context of statistical mechanics, when he wrote \cite{Gibbs} '{\em ... while the probabilities of subsequent events may often be determined from the probabilities of prior events, it is rarely the case that probabilities of prior events can be determined from those of subsequent events ...}'.  Jeffreys \cite{Jeffreys}, for the Bayesians, wrote `{\em ... a precisely stated hypothesis may attain either a high or a negligible probability as a result of observation}'.  

There is a further significant point of issue between the two approaches: frequentism requires repeatability so that an ensemble of results can be amassed, while Bayesian methods allow for the treatment of rare events, even one-offs.  Surely it is no coincidence that Fisher was an early geneticist, working in a field with plenty of data and opportunity to repeat experiments.  Jeffreys, however, was an geophysicist interested, among other things, in earthquakes which, mercifully, are rare events. A most readable account of this issue and the eventual predominance of the Bayesian view has been given by McGrayne \cite{McGrayne}.  Today it is very much the Bayesian view that holds sway \cite{Bretthorst,JaynesPapers,Lee,Box,Sivia,JaynesBook}.  It plays an essential role in information and communications theory \cite{Kullback,Cover,Goldie,MacKay} as well as in its quantum counterpart \cite{QIbook}.  Bayesian methods have been crucial in a number of scientific discoveries including the observation and interpretation of gravitational wave signals \cite{Cornish,Veitch,Littenberg,Bilby,Thrane}.

We shall proceed on the basis of the Bayesian notion of probabilities. Consider a pair of events, $A$ and $B$, each of which can take on any one of a number of possible values, which we denote by the sets $\{a_i\}$ and $\{b_j\}$.  The probability event $A$ is that associated with $a_i$ is $P(a_i)$ and similarly the probability that $B = b_j$ is $P(b_j)$.  A complete description of the two events is given by the joint probabilities $P(a_i,b_j)$, where the comma denotes ``and''.  We can recover the probabilities for the individual events from the joint probabilities by summation:
\begin{eqnarray}
P(a_i) &=& \sum_j P(a_i,b_j)  \nonumber \\
P(b_j) &=& \sum_i P(a_i,b_j) .
\end{eqnarray}

If we learn of event $A$, so that $A = a_0$ say, then the probabilities for event $B$ will change to
\begin{equation}
P(b_j|a_0) \neq P(b_j) ,
\end{equation}
where the vertical line denotes ``given'', so we read $P(b_j|a_0)$ as the probability that $B = b_j$ {\em given} that $A = a_0$.  The quantities $\{P(b_j|a_0)\}$ are conditional probabilities, the probabilities for event $B$ conditioned on knowing event $A$. The conditional probabilities are related to the joint probabilities by
\begin{eqnarray}
P(a_i,b_j) &=& P(b_j|a_i)P(a_i)  \nonumber \\
&=& P(a_i|b_j)P(b_j) .
\end{eqnarray}
These rules are an algebraic expression of the familiar probability trees we learn about in our first courses on probability theory. We can use these equations to obtain a relationship between the two sets of conditional probabilities, $A$ given $B$ and $B$ given $A$:
\begin{equation}
\label{bayes}
P(a_i|b_j) = \frac{P(b_j|a_i)P(a_i)}{P(b_j)} 
= \frac{P(b_j|a_i)P(a_i)}{\sum_k P(b_j|a_k)P(a_k)} ,
\end{equation}
which is Bayes' theorem.  We have not introduced as yet any notion of causality or time evolution, but if $A$ happens before $B$ then $P(b_j|a_i)$ is a {\em predictive} probability and $P(a_i|b_j)$ is a {\em retrodictive} probability.  The predictive conditional probabilities, $P(b_j|a_i)$, are the probabilities for the possible outcomes of event $B$ given knowledge of the earlier event $A$, $A = a_i$.  The retrodictive conditional probabilities,   $P(a_i|b_j)$, however, are the probabilities that the earlier event was $A = a_i$ given that the later event was $B = b_j$.

Our presentation of conditional probabilities and Bayes' theorem seems so natural that one may wonder what is the source of the controversy. To see this, we write Bayes' theorem in the form
\begin{equation}
P(a_i|b_j) \propto P(b_j|a_i)P(a_i) ,
\end{equation}
which we interpret as modification of the probability that $A = a_i$ on learning that $B = b_j$, but how do we interpret the initial $P(a_i)$?  A frequentist would require this to be obtained on the basis of frequency of occurrence.  For the Bayesian, however, it is just the best estimate available based on available information.
(More precisely it derives from the most uniform or unbiased probability distribution consistent with the known facts \cite{JaynesBook,QIbook}.)  This means, in particular, that different people, having access to different amounts of information, will assign different probabilities\footnote{One can sympathise with Jeffreys, who at the time was essentially a lone Bayesian, when he wrote \cite{Jeffreys} `{\em Most of the present books on statistics, and longer papers in journals, include a careful disclaimer that the authors propose to use inverse probability, and emphasise its lack of logical foundation, which is supposed to have been repeatedly pointed out.  In fact the continued mention of a principle that everybody is completely convinced is nonsense recalls the saying of the Queen in} Hamlet: {\em ``The lady doth protest too much, methinks.''}'.}. 


\section{Retrodiction in Quantum Theory}
\label{Sect3}

Our preceding presentation of conditional probabilities and Bayes' theorem may seem straightforward and, indeed, natural.  What has perhaps been insufficiently appreciated, however, is the necessary implications of this for quantum theory.  If we accept the notion that probabilities may be different for individuals having access to distinct information, as Bayesian methods demand, then it necessarily follows that in quantum theory these individuals will assign different states at any given time to the quantum system in question, as the states are used to calculate the conditional probabilities linked by Bayes' theorem.  

\begin{figure}[h]
\centering
\includegraphics[width=12cm]{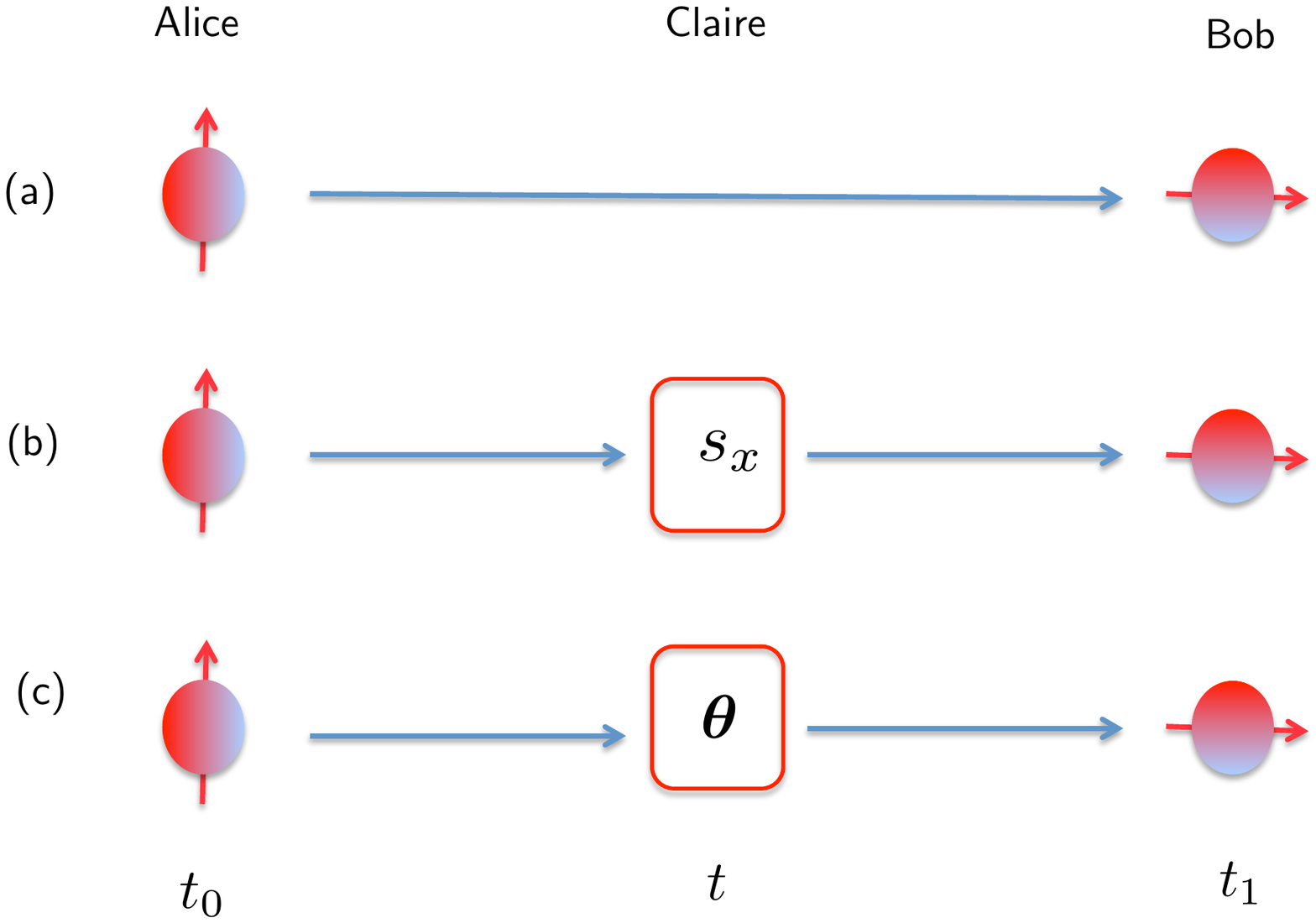}
\caption{Schematic of three sequences of spin preparation and measurements:
(\textbf{a})  Alice prepares the spin-half particle in the state $|\uparrow\rangle$ at time $t_0$ and Bob measures the $x$-component of the spin to correspond to the state $|\rightarrow\rangle$ at time $t_1$.  (\textbf{b}) Alice and Bob act as in (a) but also Claire measures $x$-component of the spin at time $t$. (\textbf{c}) Alice and Bob again act as in (a) but also Claire measures the spin along the direction $\mbox{\boldmath$\theta$}$ at time $t$.
}
\label{Fig1}
\end{figure}

A simple example, which we have presented previously \cite{SUSSP}, serves to illustrate the key idea.  Let us suppose that we have a spin-half particle experiencing a sequence of events as depicted in Fig. \ref{Fig1}.  At time $t_0$ a first individual, Alice, performs a preparation event, setting the $z$-component of the spin to $\frac{\hbar}{2}$.  At a later time, $t_1$, a second person, Bob, performs a measurement of the $x$-component of the spin and finds the value $\frac{\hbar}{2}$.  What is the state of the spin between the initial preparation event and the subsequent measurement?  The standard (predictive) quantum theory approach (the Copenhagen interpretation) would have it that the state was $|\uparrow\rangle$ and that at time $t_0$ the state collapsed into the state $|\rightarrow\rangle$.  The retrodictive approach, however, would have it that the premeasurement state is $|\rightarrow\rangle$. 

To see that the above assignment makes sense, let us suppose that a third individual, Claire, made an ideal measurement of the $x$-component of the spin at a time $t$.  What would the result of that measurement have been?  From the predictive perspective, Claire's measurement prepares the spin in an eigenstate of $s_x$, the $x$-component of the spin, and that eigenstate must be $|\rightarrow\rangle$ because of the result of Bob's later measurement.  The retrodictive picture is more direct: Bob measured the spin and found it to be the eigenvalue associated with the state $|\rightarrow\rangle$ and hence he can infer the premeasurement retrodictive state to be $|\rightarrow\rangle$.  From this the result of Claire's measurement follows directly.

For Alice, the natural description of the spin after time $t_0$ is the state $|\uparrow\rangle$.  Fully analogously, for Bob the natural state for times before $t_1$ is the state $|\rightarrow\rangle$, exploiting the information available to him.  To see the essential symmetry of this situation we consider a different measurement carried out by Claire. Let us examine what happens if Claire performs an ideal measurement of the spin along some direction $\mbox{\boldmath$\theta$}$, with results parallel and anti-parallel  to this direction corresponding to the orthogonal states $|\mbox{\boldmath$\theta$}\rangle$ and $|\mbox{\boldmath$\theta$}^\perp\rangle$ respectively. What is the probability that Claire's measurement corresponds to the state  $|\mbox{\boldmath$\theta$}\rangle$?  The answer, of course, depends on what else we know.  For Claire, who performed the measurement and knows the outcome, the probability is either $1$ or $0$.  Alice, acting alone knows only that she prepared the spin in the state $|\uparrow\rangle$ and hence calculates the probability to be
\begin{equation}
    P(\mbox{\boldmath$\theta$}|a_\uparrow) 
    = |\langle\mbox{\boldmath$\theta$}|\uparrow\rangle|^2 .
\end{equation}
If Bob knows only the result of his later measurement, then he
infers the probability 
\begin{equation}
    P(\mbox{\boldmath$\theta$}|b_\rightarrow) 
    = |\langle\mbox{\boldmath$\theta$}|\rightarrow\rangle|^2 .
\end{equation}
Both of these are in accord with the familiar Born rule.  More subtle is the form of the probability if Alice and Bob share their information.  A straightforward application of Bayes' rule gives the probability 
\begin{equation}
\label{clairespinprob}
    P(\mbox{\boldmath$\theta$}|a_\uparrow,b_\rightarrow) 
    = \frac{|\langle \rightarrow, b|\mbox{\boldmath$\theta$}\rangle|^2
    |\langle\mbox{\boldmath$\theta$}|\uparrow, a\rangle|^2}
    {|\langle \rightarrow, b|\mbox{\boldmath$\theta$}\rangle|^2
    |\langle\mbox{\boldmath$\theta$}|\uparrow, a\rangle|^2 + 
    |\langle \rightarrow, b|\mbox{\boldmath$\theta$}^\perp\rangle|^2
    |\langle\mbox{\boldmath$\theta$}^\perp|\uparrow, a\rangle|^2} .
\end{equation}
Here we have retained the labels $a$ and $b$ on the state vectors to emphasise the symmetry between them; Alice's preparation and Bob's measurement play equal, and indeed \emph{interchangeable}, roles in this probability.  It is of \emph{especial importance} for our later discussion to note that this probability does not take the form of the modulus squared overlap between a single state and a measurement operator as in the familiar Born rule.  There is nothing mysterious about this and we find a similar departure from the simple Born rule whenever we have information about both preparation and subsequent measurement events.

We note that in the denominator of equation (\ref{clairespinprob}) we are adding the probabilities of two mutually exclusive pathways from Alice measuring the spin to be in the $+z$ direction to Bob measuring it in the $+x$ direction. As equation (\ref{clairespinprob}) appears to be quantum mechanical in nature as opposed to equation (\ref{bayes}), it might be asked why we are not adding the amplitudes of the paths instead, in line with Feynman’s path integral formulation of quantum mechanics \cite{F48,F05,F10}. The reason is that here we are dealing with experimentally recordable measurement events, so the frequentist, and thus Bayesian, probability rules apply.


\section{Retrodictive states and dynamics}
\label{Sect4}

Our aim is to develop retrodictive quantum theory on a par with the more
familiar predictive theory.  For this reason it is necessary to understand
the form and properties of retrodictive states and their dynamics.  We 
start by considering a state preparation event followed by a measurement.  
Alice prepares, at time $t_0$, a quantum system in one of a complete set of 
orthonormal states $\{|i\rangle\}$.  At a later time, $t_1$, Bob makes a 
measurement of an observable with non-degenerate eigenstates $\{|m\rangle\}$.
For the present we shall assume that Alice chooses among the states 
$\{|i\rangle\}$ with equal probability.  It is of the first importance to
appreciate that this is \emph{not essential}, as we shall see, but it
does simplify the initial presentation.  Much confusion has arisen over this
point, with some authors insisting that retrodictive quantum theory applies
only for such unbiased state preparation (with the \emph{a priori}
density operator being proportional to the identity) \cite{Belinfante,Amri}.
This is not correct, as we shall show in the following sections.

Alice assigns the state on the basis of the information available to
her, so for her the post preparation state will be $|I\rangle$ say.
Bob knows only the result of his measurement and so follows the 
retrodictive approach and assigns the premeasurement state to be
$|M\rangle$, corresponding to the measurement result $M$.  The initial
predictive state is $|I\rangle$ and the final retrodictive state is
$|M\rangle$.  Between these events, the predictive state evolves 
according to the Schr\"{o}dinger equation with the formal solution
\begin{equation}
    |I(t)\rangle = \hat{U}(t,t_0)|I\rangle ,
\end{equation}
where, for a time-independent Hamiltonian, our unitary operator is
\begin{equation}
    \hat{U}(t,t_0) = \exp\left(-i\hat{H}(t-t_0)/\hbar\right) .
\end{equation}
The probability, according to Alice, that Bob's measurement
gives the result $m$ is then
\begin{equation}
    P(m|I) = |\langle m|\hat{U}(t_1,t_0)|I\rangle|^2 .
\end{equation}
For Bob, who knows only the result of his measurement, the quantity of
interest is the probability that Alice selected any given state $i$, and for
this he calculates
\begin{equation}
    P(i|M) = |\langle i|\hat{U}^\dag(t_1,t_0)|M\rangle|^2 .
\end{equation}
Note that, purely because of the simple way we have set up our model,
the two probabilities $P(M|I)$ and $P(I|M)$ are numerically equal.
The interpretations of these two conditional probabilities, however
are rather different.  It is usual and natural to understand the 
predictive conditional probability $P(M|I)$ in terms of the evolution
of the initial state, $|I\rangle$, up to time $t_1$.  In retrodictive
theory we analogously understand the conditional probability in terms 
of the evolution of the final measured state, $|M\rangle$, 
\emph{backwards} in time from $t_1$ to time $t_0$.

We can derive the dynamics of our retrodictive state from the form of 
the amplitude for finding the evolved initial state $|i\rangle$ to 
be in the state $|m\rangle$:
\begin{eqnarray}
c_{mi} &=&  \langle m|\hat{U}(t_1,t_0)|i\rangle  \nonumber \\
&=& \langle m|\hat{U}(t_1,t)\hat{U}(t,t_0)|i\rangle  \nonumber \\
&=& \langle m(t)|i(t)\rangle ,
\end{eqnarray}
where $|m(t)\rangle$ is the state corresponding to the measurement
outcome evolved backwards in time from $t_1$ to $t$.  The equation
of this backwards evolution follows directly from the fact that 
$c_{mi}$ is independent of $t$:
\begin{eqnarray}
\frac{dc_{mi}}{dt} &=& 0 \nonumber \\
&=& \frac{d}{dt}\langle m(t)| i(t)\rangle \nonumber \\
&=& \langle m(t)|\left(-\frac{i}{\hbar}\hat{H}|i(t)\rangle\right)
+ \langle \dot{m}(t)|i(t)\rangle \nonumber \\
\Rightarrow i\hbar \frac{d}{dt}|m(t)\rangle &=& \hat{H}|m(t)\rangle .
\end{eqnarray}
It follows that the retrodictive state satisfies the very same equation
as the more familiar predictive state.  There is a very significant 
difference, however: predictive states start with an initial boundary
condition and evolve forwards in time, but retrodictive states evolve
backwards in time from a final boundary condition. We might note that here we have used Schr\"{o}dinger's evolution equation but the same result is obtainable from Feynman’s path integral approach \cite{F48,F05,F10}.

Retrodictive evolution, like its predictive counterpart, is not limited
to unitary dynamics.  The same principle as presented here can be used
to derive the retrodictive evolution of mixed states associated,
for example, with imperfect final measurements.  It is also possible to
derive retrodicitve evolution for open-system dynamics such as that 
normally associated with Markovian master equations \cite{BPJJ01}.  A
simple example is the evolution associated with linear loss from an 
optical field mode.  The corresponding evolution is simply that 
associated with a linear amplifier \cite{BPJJL00}.


\section{Bayesian Inference and Quantum Retrodiction}
\label{Sect5}

It must be appreciated that retrodiction can be applied beyond the 
simple situation of unbiased state preparation presented in the preceding
section.  Here we show how the predictive and retrodictive formalisms
are linked by Bayes' theorem.  Indeed it is possible to derive the retrodictive formulation by assuming the familiar predictive quantum
theory together with Bayes' theorem \cite{BPJ00}.

It is simplest to consider the connection between retrodiction and Bayes'
theorem by reference to a quantum communications problem.  To this end let
the transmitting party, Alice, prepare and send a quantum system to a 
receiving party, Bob, who performs an ideal measurement on it and faithfully records the result.  Alice
selects one of a set of states with state vectors $\{|\phi_i\rangle\}$
with preselection probabilities $p_i$.  We impose no restriction on the states, which
need not be orthogonal nor complete, nor do we restrict the possible values
of the preparation probabilities.  It follows that the \emph{a priori}
density operator for the prepared (predictive) state is
\begin{equation}
    \hat{\rho} = \sum_i p_i |\phi_i\rangle\langle\phi_i| .
\end{equation}
Again for simplicity, we assume that the system does not evolve between
preparation and Bob's later measurement.  (Including any dynamics is 
straightforward but may obscure the argument.)  Let Bob perform an ideal 
projective measurement with outcomes $m$ corresponding to the non-degenerate
eigenvectors $|m\rangle$, so that each measurement outcome corresponds
to a single state vector.  Neither restricting Alice to preparing 
pure states nor Bob to ideal projective measurements is necessary, but
does simplify the mathematical presentation.  A fully general treatment in
which Alice prepares mixed states and Bob performs generalized measurements
can be found in \cite{BPJ00}.  A brief account of this is presented in
Appendix \ref{AppA}.

We can follow our discussion of predictive and retrodictive probabilities
from section 2 and construct the relevant probabilities.  The probability
that Alice prepares the state $|\phi_i\rangle$ and that Bob measures
and records the outcome $m$ is
\begin{equation}
    P(a_i,b_m) = p_i|\langle m|\phi_i\rangle|^2 .
\end{equation}
From this we find the probabilities for Alice's preparation event and Bob's
measurement outcome:
\begin{equation}
    P(a_i) = \sum_m p_i|\langle m|\phi_i\rangle|^2 = p_i ,
\end{equation}
as required, and 
\begin{equation}
    P(b_m) = \sum_i p_i|\langle m|\phi_i\rangle|^2 = 
    \langle m|\hat{\rho}|m\rangle .
\end{equation}
We note that Alice's preparation probabilties are \emph{independent} of
Bob's measurement outcomes but that the \emph{a priori} probabilities
of Bob's measurement outcomes will, in general, depend on Alice's ensemble
of states. Only for the unbiased case in which the \emph{a priori} 
density operator is proportional to the identity, $\hat{\rho} \propto
\hat{\rm I}$, will the probabilities for each of Bob's possible recorded measurement
results be equal and, in this sense, independent of Alice's ensemble of possible
preparation events.

Bayes' rule enables us to calculate the conditional probabilities 
for Bob's measurement results and for Alice's preparation events:
\begin{eqnarray} \label{Eq19}
P(b_m|a_i) &=& |\langle m|\phi_i\rangle|^2 \\
\label{Eq20}
P(a_i|b_m) &=& \frac{p_i|\langle m|\phi_i\rangle|^2}
{\sum_j p_j|\langle m|\phi_j\rangle|^2} .
\end{eqnarray}
Note that in general the conditional probability $P(a_i|b_m)$, unlike
$P(b_m|a_i)$, is not
the simple overlap between a pair of states, one for the preparation
event and the other associated with the measurement outcome.  There is
nothing mysterious about this, it is simply a consequence of the fact that
Bob has prior information about Alice's preparation event, in that the
\emph{a priori} density operator is not the fully mixed state, proportional
to the identity.  On the other hand, we have assumed ideal measurement and 
faithful recording, which means equal {\em a priori} postselection probabilities
in the recording of results.  Biased postselection might arise 
through instrumental imperfection or as a deliberate choice by Bob to discard 
some of the measurement results, as often happens, for example, in quantum optics experiments when photons are not detected and the experimental run is discarded.  
In this case the predictive probability $P(b_m|a_i)$ would be {\em symmetric}
in form with the retrodictive probability $P(a_i|b_m)$ with {\em a priori}
postselection probabilities $\bar{p}_m$ instead of the {\em a priori} 
preselection probabilities $p_j$:
\begin{equation}
\label{Eq21}
    P(b_m|a_i) = \frac{\bar{p}_m|\langle m|\phi_i\rangle|^2}
    {\sum_n\bar{p}_n|\langle n|\phi_i\rangle|^2} .
\end{equation}
Both the predictive and the retrodictive conditional probabilities depend on
the overlap of the prepared (predictive) states $|\phi_i\rangle$ and the 
recorded measured (retrodictive) states $|m\rangle$.



\section{Controversies, Objections and Resolutions}
\label{Sect6}

The introduction of quantum retrodiction, rather like Bayes' 
theorem \cite{McGrayne}, has been beset with objections and 
controversies.  We explore the more visible of these here and
assess why they arose and uncover the subtle problems with them.
There are three principal objections that have been raised to the 
general application of quantum retrodiction as we have presented it
here.  These are (i) that the wavefunction or state vector has a real
existence and that it evolves forwards in time, (ii) that quantum theory 
applies only to large ensembles, (iii) that quantum
retrodiction is valid but only for unbiased initial states (with the
\emph{a priori} density operator proportional to the identity) and
finally (iv) that retrodiction is valid but that in order to arrive at a
time-symmetric formulation, the retrodictive state 
must depend on \emph{both} the final measurement and also the initial
density operator.  We address each of these in turn.

\subsection{Reality of the Wavefunction?}

It is the very essence of the Bayesian inference that probabilities 
depend on one's knowledge and that acquiring information enables one
to modify these.  In constructing retrodictive quantum theory, we 
exploited this feature to allow both Alice and Bob to assign a state
to a quantum system between the preparation and measurement events,
based on the information available to them.  Key to our ability to do 
this is the Bayesian idea that the probabilities for the quantum
system between the preparation and measurement events depend on whether
the individual has information from before or after any observation
occurring between the preparation and later measurement events.  In
general the predictive state assigned by Alice and the retrodictive 
states assigned by Bob will be different but, as we have seen, either
or both of these states can be used to calculate correctly the probability
for any intervening measurement event.

Whether one holds to the idea of the physical reality a wavefunction or 
state vector based on a preparation event, a calculation by Bob based
on a rival retrodictive state produces the correct probabilities and
this suggests that the retrodictive state has as much validity and 
reality as the more familiar predictive state.  To insist that
the predictive state is real and that it must be used, if applied
in conjunction with Bayes' rule, will generate the correct retrodictive
probabilities but precisely the same retrodictive probabilities can
be obtained more directly by use of the retrodictive state.  It seems to
us that the situation for one taking this position is in much the same
position as the early statisticians whom Jeffreys challenged.  To be clear,
one can take the view that the (predictive) wavefunction is real and must
be used, but is then left in the somewhat awkward position of having to
explain why calculations based on the retrodictive state give the very 
same probabilities as those recovered from a predictive state together with
Bayes' theorem.

The insistence on a real forward-propagating wavefunction can lead to the well-known Einstein-Podolsky-Rosen paradox. The retrodictive approach sheds an interesting light on experiments such as that of Kocher and Commins \cite{KC67} that involve this paradox. Here a three-level atom emits two photons in opposite directions with correlated polarisations. In the usual predictive picture, the polarised detection of one photon collapses the combined two-photon state, instantaneously affecting the polarisation of the other photon even though this can be a large distance away. In the retrodictive approach, however, the collapse of the retrodictive state from the detection of the first photon emitted can occur at the atom itself at the time of emission and can thus determine the correlated intermediate atomic state, which then determines the polarisation of the second photon, removing the paradox \cite{PB99}. It is interesting to compare this with the Wheeler-Feynman absorber theory \cite{WF45}. Here the advanced field sent backwards in time from the polarised detector upon detection of the first emitted photon causes the atom to jump to a particular intermediate state at the time of emission, resulting in the later emission of the second retarded field with correlated polarisation, again removing the paradox \cite{P82}.

\subsection{Quantum Theory Applies only to Ensembles?}

The probabilistic interpretation of quantum theory appears in every
introductory textbook on the subject, but explicit mention of how 
the probabilities are to be calculated receives less attention.  Schiff,
in his famous text, devotes a short section to the statistical 
interpretation in which it is clear that he is assigning probabilities
on the basis of the number of outcomes for each measured value, carried
out on a large number or copies \cite{Schiff}.  Peres, in his text, is
more explicit when he writes '\emph{Here a probability is defined as
usual: If we repeat the same preparation many times, the probability 
of a given outcome is its relative frequency, namely the limit of the
ratio of the number of occurrences of that outcome to the total number of
trials, when these numbers tend to infinity.}' \cite{PeresBook}.  These
statements are very much in the spirit of the frequentist interpretation
of probability and therefore of quantum theory.  Belinfante, in his
text, emphasises the same point, that quantum theory is applicable only
to ensembles \cite{Belinfante}.  Schr\"{o}dinger famously, and picturesquely
made the same point when he wrote '\emph{... we never experiment with just 
one electron or atom or (small) molecule.  In thought experiments we
sometimes assume that we do; this invariably leads to ridiculous
consequences ...}' and, to emphasise the point, '\emph{... it is fair
to state that we are not experimenting with single particles, any more
than we can raise Ichthyosauria in the zoo.}' \cite{Schrodinger}.

At the times that Schr\"{o}dinger and Belinfante were writing, a restriction
on quantum theory to apply only to ensembles was perhaps justified by the
then experimental state of the art.  Subsequent technical advances, however,
have dramatically changed this position.  Today intricate experiments with
single atoms, photons, ions and other quantum systems, while not exactly
commonplace, are reported regularly and pass almost without comment
\cite{HarocheBook,Haroche,Wineland}.  Describing these experiments has 
required the application of quantum theory to single quantum systems and
new techniques involving conditional evolution, very much in the Bayesian
spirit, have been devised and widely applied.  In particular, the development
of quantum Monte-Carlo methods to model the evolution of open and also of
monitored quantum systems has enabled the modelling of single quantum 
systems \cite{Carmichael,Dalibard,Percival,Plenio,Nienhuis,Milburn}.

To withhold the possibility of attributing a wavefunction or state vector
to a single quantum system today is to deny the possibility of describing 
many current experiments and few research workers can accept that.

\subsection{Restriction to Unbiased State Preparation?}

The objections discussed in the preceding two subsections had something
of a philosophical flavour and challenged the validity of quantum
retrodiction.  The remaining objections discussed here and in the 
following subsection do not question the validity of quantum retrodiction
but rather its generality and the form of the retrodictive state.  Each 
is based on the lack of symmetry between the forms of the quantum 
predictive and retrodictive probabilities. 

The central point is that the two conditional probabilities given in 
equations (\ref{Eq19}) and (\ref{Eq20}) have different forms.  We have seen
that this is due to the prior information about the state preparation.  
If no prior information is available, then Bob assigns an initial density
operator proportional to the identity operator.  For the simple case
treated in section \ref{Sect5} this corresponds to Alice choosing from
a complete set of orthogonal states $\{|\phi_i\rangle\}$ with equal 
probabilities, so that the {\em a priori} density operator is
\begin{equation}
    \hat{\rho} = \sum_ip_i|\phi_i\rangle\langle\phi_i| =
    \frac{\hat{\rm I}}{D},
\end{equation}
where $\hat{\rm I}$ is the identity operator and $D$ is the dimension
of the state space spanned by the prepared states\footnote{More generally,
an unbiased preparation is any in which the {\em a priori} density operator
is proportional to the identity operator but, for simplicity of 
presentation, we refer the reader to \cite{BPJ00} for a treatment of the 
general case.}.  State preparation of this form is called  {\em unbiased} 
\cite{BPJ00,SUSSP}, {\em uniform} \cite{Watanabe} or {\em garbled}
\cite{Belinfante} and in this case the conditional probability $P(a_i|b_m)$
simplifies to 
\begin{equation}
    P(a_i|b_m) = |\langle m|\phi_i\rangle|^2 .
\end{equation}
This has the same form (and in this simple case also the same value) as the
predictive conditional probability $P(b_m|a_i)$, given in equation
(\ref{Eq19}).  This expression, moreover is of the form suggested by Born's
rule and it seems that this similarity is the motivation for claiming a 
restriction to the validity of quantum retrodiction \cite{Watanabe,Belinfante,Amri}.
Indeed Amri {\em et al} justify this by appealing to Gleason's theorem
\cite{Gleason,Busch} to suggest that only probabilities of this form are 
valid in quantum theory.  Were this indeed the case then it would represent
a serious challenge to the application of Bayesian inference in quantum 
theory.  We now know, however, that a simple generalisation of Gleason's
theorem shows that conditional probabilities of the form given in equation
(\ref{Eq20}) are indeed consistent with Gleason's theorem and, therefore,
with quantum theory \cite{BCJP14}.  

If we are to accept, as is our premise, that Bayes' rule applies to 
quantum probabilities just as it does to classical ones then, perforce,
we are required also to accept the form of the retrodictive conditional 
probability given in equation (\ref{Eq20}).  Clearly there is a lack
of symmetry between the simple forms of the predictive and retrodictive 
conditional probabilities, but this is a consequence of the availability
of prior information about the state preparation and the lack of any
such information concerning the outcome of the later measurement.  Neither 
Gleason's theorem nor aesthetic considerations, such as a preference for
a symmetry of form between the predictive and retrodictive conditional 
probabilities, suffice to invalidate the
application of quantum retrodiction when prior information concerning
the state preparation lead to a biased {\em a priori} state.

\subsection{Should there be a Time-Symmetric Formulation?}

Our final objection also relates to the form of the retrodictive conditional
probability, but is more subtle in that the point of issue is not the 
numerical value of this probability but rather the form of the retrodictive
state used to calculate it \cite{Fields}.  Fields {\em et al} \cite{Fields}
do insist, moreover, that 
retrodictive quantum theory, as we have presented it here, applies only
to unbiased state-preparation, even though they do not dispute the Bayesian
forms of and relationship between the predictive and retrodictive conditional
probabilities\footnote{Something of Jeffrey's frustrations with the 
frequentists, alluded to earlier, come to mind.}.  This is clearly wrong!

In order to appreciate fully
the issues, it is worth working in more generality than we have allowed
ourselves thus far.  To this end, let Alice prepare a quantum system in
one of a set of states (possibly mixed) corresponding to the density 
operators $\{\hat{\rho}_i\}$, with corresponding probabilities $\{p_i\}$, 
so that the {\em a priori} density operator is
\begin{equation}
    \hat{\rho} = \sum_ip_i\hat{\rho}_i .
\end{equation}
Further, let Bob perform a generalized measurement, with outcomes $\{m\}$
corresponding to the measurement operators $\{\hat{\pi}_m\}$, as in 
Appendix \ref{AppA}.  Our predictive and retrodictive probabilities are
then
\begin{eqnarray}
\label{Eq25}
P(b_m|a_i) &=& {\rm Tr}\left(\hat{\rho}_i\hat{\pi}_j\right) \\
\label{Eq26}
P(a_i|b_m) &=& \frac{p_i{\rm Tr}\left(\hat{\rho}_i\hat{\pi}_m\right)}
{{\rm Tr}\left(\hat{\rho}\hat{\pi}_m\right)} ,
\end{eqnarray}
respectively.  Fields {\em et al} seek to rewrite the second of these in
a similar form to the first by choosing retrodictive states and retrodictive
``measurement'' operators of the form:
\begin{eqnarray}
\hat{\rho}_m^{\rm ret,FSB} &=&  \frac{\hat{\rho}^{\frac{1}{2}}\hat{\pi}_m\hat{\rho}^{\frac{1}{2}}}
{{\rm Tr}(\hat{\rho}\hat{\pi}_m)} \\
\hat{\pi}_i^{\rm ret,FSB} &=& \hat{\rho}^{-\frac{1}{2}}p_i\hat{\rho}_i\hat{\rho}^{-\frac{1}{2}} ,
\end{eqnarray}
a transformation reminiscent of that associated with so-called square-root measurements 
\cite{QIbook,Sarah}.  Note that $\hat{\rho}_m^{\rm ret,FSB}$ has unit trace and the probability
operators $\hat{\pi}_i^{\rm ret,FSB}$ sum to the identity.  From these
it follows that the retrodictive conditional probability becomes
\begin{equation}
    P(a_i|b_m) = {\rm Tr}(\hat{\rho}_m^{\rm ret,FSB}
    \hat{\pi}_i^{\rm ret,FSB}) .
\end{equation}
Superficially, at least, it looks as though the symmetry of form between the
predictive and retrodictive probabilities has been restored.  Yet this is
not the case.  The first and most important point to appreciate is that
in the formulation of Fields {\em et al} the retrodictive state depends on
{\em both} the final and also the {\em initial state}, while the predictive
state depends only on the initial preparation event.  In this crucial sense 
their identification of $\hat{\rho}_m^{\rm ret,FSB}$ is far from being
symmetric.  A simple example serves to highlight this problem.  Let us
consider, once again, the situation depicted in figure \ref{Fig1} {\textbf{a}}.
Alice prepares a spin-half particle in the state $|\uparrow\rangle$ and,
subsequently, Bob performs a measurement with the result corresponding to 
the state $|\rightarrow\rangle$.  In this simplest of cases, the predictive
state is $|\uparrow\rangle$, while the retrodictive state is $|\rightarrow\rangle$.  For Fields {\em et al}, however, while the predictive
state remains $|\uparrow\rangle$, the retrodictive state is {\em also} 
$|\uparrow\rangle$.  Far from representing or depending on the final 
measurement outcome, the retrodictive state in this case, according to 
Fields {\em et al} is {\em independent} of it!  By insisting on the
wholly unnecessary requirement to write both predictive and retrodictive
probabilities in the same form, Fields {\em et al} have arrived at a 
situation in which the predictive state depends only on the initial 
preparation but the retrodictive state depends on both the final measurement
{\em and} the initial preparation.

Expression (\ref{Eq21}) for prediction with biased postselection can also be 
generalized for mixed states and generalized measurements.  It is interesting
to note that the resulting expression can also be written in the deceptively
simple form of the trace of the product of a unit-trace operator and an 
element of a positive operator-valued measure.  The unit-trace operator, however,
is {\em not} the prepared density operator but depends also on the later
measurement procedure.

The asymmetry between the forms of the two conditional probabilities,
$P(b_m|a_i)$ and $P(a_i|b_m)$, in equations (\ref{Eq25}) and (\ref{Eq26}), 
originates simply from our ability to 
control (at least to some extent) future events but not past ones and,
with this, to have (some) knowledge of the past but not of the future. This is a feature common to both the classical and the quantum descriptions of our world.


\section{Retrodiction, Time-Reversal and the Arrow of Time}
\label{Sect7}

In this final section we examine the implications of quantum retrodiction
for the nature of time.  We start with the observation that retrodiction
is not time-reversal, but rather should be thought of as a switch from an 
initial boundary condition to a final one.  It is perhaps surprising that 
quantum retrodiction has implications, also, for mechanisms that have
been suggested to account for or are associated with the arrow of time.

\subsection{Retrodiction is not time-reversal}

In common with most of our microscopic laws, quantum theory does not have
an explicit direction of time built in and, indeed, is symmetric under
time-reversal.  Indeed, time reversal can be associated with an explicit
anti-unitary transformation \cite{Wignerbook,Bigi}.  It is important to 
appreciate, however, that retrodiction is not time-reversal.  The simplest 
way to see this is to note that time reversal would take the evolved
quantum state at some time $t_1$ and reverse its evolution back to the
initial state at the earlier time $t_0$.  Quantum retrodiction, however,
starts with the result of a measurement carried out at time $t_1$ and
assigns a state on the basis of the measurement result and evolves
this state back towards the earlier time $t_0$.  Only in very special 
cases will the (predictive) state evolved forwards in time from $t_0$
to $t_1$ coincide with the state associated with the measurement performed 
at that time.  

Something similar happens, also, with statistical physics in the classical
domain.  As a simple example, consider the motion of a small particle
undergoing Brownian motion \cite{Mazo,Lemons}.  Let us suppose that 
this particle is prepared at position ${\bf r}_0$ at time $t_0$  and, at a 
later time time $t_1$, we observe it to be at some position ${\bf r}_1$.  
Time-reversal would mean retracing the trajectory of particle from 
${\bf r}_1$ back to its initial position ${\bf r}_0$.  Retrodiction, however,
would correspond to reversed time evolution under the same random forces
that produced the initial motion, of which we have only statistical 
knowledge.  The retrodictive analysis will produce a distribution of 
positions from which the particle might have begun its motion, and 
this distribution will be very much like that which would be produced
by a predictive analysis of the position of the particle at time $t_1$
based only on the knowledge that at it was at position ${\bf r}_0$ at
time $t_0$.

\subsection{Quantum arrow of time?}

The study of quantum retrodiction brings fresh insights into the possibility 
of a quantum arrow of time.  We have noted that at the microscopic level,
and in common with most of the rest of physics, there is no evidence of a
preferred arrow of time.  It was suggested, however, early in the development
of quantum theory that measurement might provide this direction.  In his
famous book \cite{vonNeumann}, von Neumann introduced two interventions by 
which a quantum state could evolve, one corresponding to the uncontrollable
changes that occur in a measurement (effectively what was to become known as
wavefunction collapse) and the second being the unitary evolution associated
with Schr\"{o}dinger evolution.  Bohm was yet more explicit in linking 
quantum measurement with the irreversibility associated with the direction
of time, when he wrote \cite{Bohm} '{\em Because the irreversible behavior
of the measuring apparatus is essential for the destruction of definite
phase relations and because, in turn, the destruction of phase relations
is essential for the consistency of the quantum theory as a whole, it 
follows that thermodynamic irreversibility enters into the quantum theory 
in an integral way.}'.  This description fits well with a view in which the 
wavefunction or state vector has an ontological meaning, as does the 
many-worlds interpretation of Everett \cite{Everett}, in which there is
no such dephasing but rather a branching of the universal state vector into
ever more entangled universes as time progresses \cite{Zeh}.

Retrodictive quantum theory requires us to adopt a different physical picture,
one in which the measurement process loses its privileged role.  To see this,
let us return to the simple situation discussed in section \ref{Sect4}.  There 
we evaluated the overlap between a forward-evolved initial state $|i(t)\rangle$
and the backward-evolved retrodictive state $|m(t)\rangle$.  We can identify
the time $t$ with a collapse of the state, but we
are free to choose the time $t$ to take any value between $t_0$ and $t_1$.
This freedom argues against a physical process being responsible for something
like a wavefunction collapse.  Moreover, if we prefer the many-worlds 
interpretation then we can equally well consider the retrodictive state 
splitting into multiple universes as it evolves backwards in time.  The
Copenhagen interpretation, with its wavefunction collapses, and the many-worlds
interpretation, as well as proposed modifications of quantum theory introducing
collapse mechanism or spontaneous localization \cite{GRW}, are at odds with
retrodictive reasoning, which has at its core the lack of a preferred arrow 
of time, reflecting the time-symmetric nature of the Schr\"{o}dinger equation.
These ideas are similarly at odds with Bayesian ideas in which the probability,
and hence the quantum state, are a manifestation of one's belief dependent on the
available knowledge, rather as opposed to a real or ontological existence subject
to physical influences.

Retrodictive and Bayesian methods seem to require us to reject a role for
quantum measurement (with or without wavefunction collapse) in the origin
of the arrow of time.  Indeed rather than the arrow of time originating from
quantum theory, a causal time arrow is inserted {\em into} the theory.  A
generalisation of Gleason's theorem \cite{BCJP14} shows that the probability
of a measurement outcome, without post-selection, is proportional to the trace
of the product of an operator representing the measurement outcome and an 
operator associated with the prepared state.  Imposition of the causal 
requirement that the choice of measurement apparatus cannot influence the 
probability that a particular state is prepared leads to the usual predictive
expression (\ref{EqA5}) with the preparation and measurement operators 
being a density operator and the element of a positive operator-valued 
measure respectively \cite{BCJP14,Pegg06}.
This means that we must look elsewhere for the origin of the arrow of time.
The increase of entropy, associated with the second law of thermodynamics is 
often linked with the direction of time \cite{Zeh}, but is it the origin of 
the arrow or simply a consequence of the arrow of time?
In other words, given a preferred direction of time, entropy would naturally increase in this direction, rather as the pattern of leaves blown from a tree are a consequence of the direction of the wind but are not the 
origin of this direction \cite{VaccaroP}.  Perhaps the most satisfactory idea at present, from the perspective of quantum retrodiction, is that proposed by Vaccaro \cite{Vaccaro1,Vaccaro2}, in which the arrow of time can be traced back to small differences in the forward and backward Hamiltonians arising from T-violating interactions.

\section{Conclusions}
\label{Sect8}

Quantum retrodiction allows one to assign a state to a quantum system 
prior to observation, based solely on the result of the measurement.  This
retrodictive state can be evolved backwards in time to make probablistic 
statements about earlier events, including the initial preparation event.
We have seen that the link between retrodictive quantum mechanics and the
more familiar predictive form follows as a natural consequence of Bayes'
theorem and that this brings with it, naturally, the Bayesian interpretation of quantum probabilities.

We have presented and addressed the principal objections that have been
raised either against quantum retrodiction or have suggested limits to its
validity.  The key to this is clearly the adoption of a Bayesian interpretation of quantum probabilities.  Indeed it is possible to derive retrodictive quantum mechanics from the conventional predictive quantum theory, coupled with Bayesian inference \cite{BPJ00}.

Adopting quantum retrodiction has implications for the interpretation of the wavefunction which, in common with Bayesian probabilities is more a statement of knowledge rather than something with a physical existence.  This has implications also for the origins of the arrow of time.  In particular, quantum retrodiction suggests that any state-collapse associated with a measurement can be thought of as occurring at any time between the preparation and measurement events, and it is difficult to reconcile this freedom with a physical change to the wavefunction.

\section*{Author Contributions}All of the authors contributed to the research described in this paper.  SMB drafted the original manuscript, which was amended and edited following comments from JJ and DTP.
\begin{acknowledgments}
This work was supported by the Royal Society, grant number 
RP150122.  DTP thanks D. Matthew Pegg for his support.
\end{acknowledgments}
\vfill

\appendix
\section{Retrodiction for Mixed States and Generalized Measurements}
\label{AppA}

In Section \ref{Sect5} we introduced, for simplicity, the idea of a 
retrodictive state by reference only to the preparation of pure states
followed by ideal projective measurements.  We show here that quantum
retrodiction can be applied more generally, when mixed states are
prepared, followed by generalized measurements \cite{QIbook,Sarah}.
We follow the same line of reasoning as that in Section \ref{Sect5}.

Let Alice choose, with probabilities $p_i$, from among a set of mixed
states with density operators $\hat{\rho}_i$.  It follows that the
\emph{a priori} density operator for the prepared (predictive) state is
\begin{equation}
    \hat{\rho} = \sum_i p_i\hat{\rho}_i .
\end{equation}
As in Section \ref{Sect5} we assume, for simplicity, that the states
$\hat{\rho}_i$ do not evolve between preparation and measurement.  Bob
then performs a generalized measurement described by a probability
operator measure (POM) or positive operator-valued measure (POVM) with
the outcome $m$ associated with the positive operator $\hat{\pi}_m$,
so that the probability that Alice prepares the state $\hat{\rho}_i$
and that Bob finds the measurement result $m$ is
\begin{equation}
    P(a_i,b_m) = p_i{\rm Tr}\left(\hat{\rho}_i\hat{\pi}_m\right) .
\end{equation}
From this we find the preparation and measurement probabilities 
\begin{eqnarray}
P(a_i) &=& \sum_m p_i{\rm Tr}\left(\hat{\rho}_i\hat{\pi}_m\right)
= p_i \\
P(b_m) &=& \sum_i p_i{\rm Tr}\left(\hat{\rho}_i\hat{\pi}_m\right)
= {\rm Tr}\left(\hat{\rho}\hat{\pi}_m\right) .
\end{eqnarray}
Bayes' rule enables us to calculate the conditional probabilities:
\begin{eqnarray}
\label{EqA5}
P(b_m|a_i) &=& {\rm Tr}\left(\hat{\rho}_i\hat{\pi}_j\right) \\
P(a_i|b_m) &=& \frac{p_i{\rm Tr}\left(\hat{\rho}_i\hat{\pi}_m\right)}
{{\rm Tr}\left(\hat{\rho}\hat{\pi}_m\right)} .
\end{eqnarray}
Note that the predictive probability, $P(b_m|a_i)$, is simply the 
trace of the product of two positive operators, $\hat{\rho}_i$ and
$\hat{\pi}_m$.  The retrodictive probability, $P(a_i|b_m)$, however,
has a more complicated form.

In the simple analysis presented in Section \ref{Sect5} the retrodictive
state associated with the measurement result $m$ was simply that 
associated with the pure-state ket $|m\rangle$.  For generalized measurements, with
the outcome $m$ associated with the probability operator (POVM element)
$\hat{\pi}_m$, the retrodictive state is rather a mixed state with 
density operator
\begin{equation}
    \hat{\rho}_m^{\rm retr} = \frac{\hat{\pi}_m}
    {{\rm Tr}\left(\hat{\pi}_m\right)} .
\end{equation}
The denominator ensures that the trace of $\hat{\rho}_m^{\rm retr}$ is
unity, as required for any density operator.  This is the natural assignment given that we wish the retrodictive state to depend only on 
the measurement outcome and not on the earlier state preparation.  For
an ideal projective measurement, our probability operator reduces to
$\hat{\pi}_m = |m\rangle\langle m|$ and we recover the retrodictive state
identified in Section \ref{Sect5}. 
\end{document}